\documentclass[12pt]{article}
\newcommand{\ben}{\begin{enumerate}}
\newcommand{\een}{\end{enumerate}}
\newcommand{\beq}{\begin{equation}}
\newcommand{\eeq}{\end{equation}}
\newcommand{\bse}{\begin{subequation}}
\newcommand{\ese}{\end{subequation}}
\newcommand{\bea}{\begin{eqnarray}}
\newcommand{\eea}{\end{eqnarray}}
\newcommand{\bc}{\begin{center}}
\newcommand{\ec}{\end{center}}

\def\DR{\rm I\kern-1.45pt\rm R}
\def\DC{\kern2pt {\hbox{\sqi I}}\kern-4.2pt\rm C}
\def\DH{\rm I\kern-1.5pt\rm H\kern-1.5pt\rm I}
\begin{document}

\begin{center}
{\Large\bf 4D singular oscillator and  \\  [2mm] generalized
MIC-Kepler system}\\[3mm]
{\bf L.G. Mardoyan and M.G. Petrosyan} \\[3mm]
Artsakh State University, Stepanakert and Yerevan State University, Yerevan, Armenia\\
1, Alex Manoogian st., 375025, Yerevan, Armenia; e-mail:
mardoyan@ysu.am
\end{center}

\vspace{3mm}
\begin{abstract}
It is shown that the generalized MIC-Kepler system and
four-dimensional singular oscillator are dual to each other and
the duality transformation is the generalized version of the
Kustaanheimo-Stiefel transformation.
\end{abstract}

\vspace{0.5cm}

The Schr\"{o}dinger equation for a generalized MIC-Kepler or
charge-dyon system has the form \cite{mard1}
\bea \frac{1}{2\mu}\left({\hat p}_i + \frac{e}{c}
A_i^{(\pm)}\right)^2\,\psi^{(\pm)} +\left[\frac{\hbar^2 s^2}{2\mu
r^2}-\frac{e^2}{r}+\frac{\lambda_1}{r(r+z)} +
\frac{\lambda_2}{r(r-z)}\right]\psi^{(\pm)}=E\psi^{(\pm)},
\label{schr} \eea
where $\lambda_{1}$ and $\lambda_{2}$ are nonnegative constants.
We recall that a dyon is a hypothetical particle introduced by
Schwinger \cite{Schwinger} and is a source of both an electric and
magnetic field. The vector potentials
\begin{eqnarray*}
{\bf A}^{(\pm)}=\frac{1}{r(r \mp z)}(\pm y, \mp x,0)
\end{eqnarray*}
correspond to a Dirac monopole \cite{Dirac} with the magnetic
charge $g=\hbar cs/e$ $(s=0,\pm1/2,\pm 1,\ldots)$ and with the
axes $z > 0$ and $z < 0$ correspondly. It is easily seen that the
vector potentials $A_i^{(+)}$ and $A_i^{(-)}$ are connected by a
gauge transformation
\begin{eqnarray*}
A_i^{(-)}=A_i^{(+)}+\frac{\partial f}{\partial x_i},
\end{eqnarray*}
where $f= 2g\arctan (y/x)$ and the strength of the dyon magnetic
field is
\begin{eqnarray*}
{\bf B} = {\bf \nabla} \times {\bf A}^{(\pm)} = g\frac{\bf
r}{r^3}.
\end{eqnarray*}

It should be noted that the Schr\"{o}dinger equation (\ref{schr})
for $\lambda_1=\lambda_2=0$ and $s=0$ reduces to the
Schr\"{o}dinger equation of the MIC-Kepler system \cite{Z, mic}.
At $s=0$ Eq.~(\ref{schr}) is reduced to the Schr\"{o}dinger
equation for the generalized Kepler-Coulomb problem \cite{KMP1}.
In case when $s=0$ and $c_1=c_2\neq 0$, the equation (\ref{schr})
reduces to the Hartmann system that has been used for describing
axially symmetric systems like ring-shaped molecules~\cite{hart}.

In~\cite{mard1,mard2} it is shown that the variables in the
Schr\"{o}dinger equation (\ref{schr}) are separated in spherical,
parabolic and prolate spheroidal coordinates. For completeness,
here we present the explicit forms of the spherical and parabolic
bases of the generalized MIC-Kepler system found in~\cite{mard1}.

In the spherical coordinates
\bea x=r\sin\theta\cos\varphi, \qquad y=r\sin\theta\sin\varphi,
\qquad z =r\cos\theta \label{spherical} \eea
the wave function of the generalized MIC-Kepler system has the
form
\bea \psi \equiv
\psi_{njm}^{(s)}\left(r,\theta,\varphi;\delta_{1},\delta_{2}\right)=
R_{nj}^{(s)}\left(r;\delta_{1},\delta_{2}\right)\,
Z_{jm}^{(s)}\left(\theta,\varphi;\delta_{1},\delta_{2}\right).
\label{sbasis} \eea
The functions
$Z_{jm}^{(s)}\left(\theta,\varphi;\delta_{1},\delta_{2}\right)$
and $R_{nj}^{(s)}\left(r;\delta_{1},\delta_{2}\right)$ are given
by the formulae
\begin{eqnarray*} Z_{jm}^{(s)}(\theta, \varphi; \delta_{1}, \delta_{2}
)=N_{jm}(\delta_{1},
\delta_{2})\left(\cos\frac{\theta}{2}\right)^{m_1}
\left(\sin\frac{\theta}{2}\right)^{m_2}
P_{j-m_+}^{(m_2,m_1)}(\cos\theta) e^{i(m+s)\varphi},
\end{eqnarray*}
\begin{eqnarray*} R_{nj}^{(s)}(r)= C_{nj}(\delta_1, \delta_2)(2\varepsilon
r)^{j+\frac{\delta_1+ \delta_2}{2}}e^{-\varepsilon
r}F\left(-n+j+1; 2j+\delta_1+ \delta_2+2; 2\varepsilon r\right),
\end{eqnarray*}
where $P_{n}^{(\alpha,\beta)}(x)$ are the Jacobi polynomials,
$F(a; c; x)$ is the confluent hypergeometric function,
$N_{jm}(\delta_{1}, \delta_{2})$ and $C_{nj}(\delta_1, \delta_2)$
are normalization constants
\begin{eqnarray*} N_{jm}(\delta_{1},
\delta_{2})=\sqrt{\frac{(2j+\delta_{1}+\delta_{2}+1)(j-m_+)!
\Gamma(j+m_+ +\delta_{1}+\delta_{2}+1)}{4\pi\Gamma(j-m_-
+\delta_{1}+1) \Gamma(j+m_- +\delta_{2}+1)}}, \end{eqnarray*}
\begin{eqnarray*} C_{nj}(\delta_1, \delta_2)=
\frac{2\varepsilon^{2}}{\Gamma\left(2j+\delta_1+
\delta_2+2\right)}\sqrt{\frac{\Gamma\left(n+j+\delta_1+
\delta_2+1\right)}{(n-j-1)!}}. \end{eqnarray*}
We denote the following expression by $\varepsilon$:
\begin{eqnarray*} \varepsilon= \sqrt{-\frac{2\mu E}{\hbar^2}} =
\frac{1}{a\left(n+\frac{\delta_1+ \delta_2}{2}\right)},
\end{eqnarray*}
where $a=\hbar^2/\mu e^2$ is the Bohr radius. The energy spectrum
has the form
\bea E \equiv E_n^{(s)} = -\frac{\mu
e^4}{2\hbar^2\left(n+\frac{\delta_1+ \delta_2}{2}\right)^{2}}
\label{energy}\eea
and the quantum numbers $m$ and $j$ run through the values:
$m=-j,-j+1,\dots,j-1,j$ and $j=m_+, m_+ +1,\dots, n-1$. We also
make the following notation:
\begin{eqnarray*}
m_{\pm} = \frac{|m+s| \pm |m-s|}{2}, \qquad m_{1,2} = |m \pm s|+
\delta_{1,2} = \sqrt{(m \pm s)^2 +\frac{4\mu
\lambda_{1,2}}{\hbar^2}}.
\end{eqnarray*}
The wave functions (\ref{sbasis}) are the eigenfunctions of
commuting operators ${\hat M}$ and ${\hat J}_z$ and
\begin{eqnarray*}
 \hat{M}\psi_{njm}^{(s)}(r,\theta,\varphi;\delta_{1},\delta_{2})=
\left(j+\frac{\delta_{1}+\delta_{2}}{2}\right)
\left(j+\frac{\delta_{1}+\delta_{2}}{2}+1\right)
\psi_{njm}^{(s)}(r,\theta,\varphi;\delta_{1},\delta_{2}),
\end{eqnarray*}
where
\begin{eqnarray*} \hat{M}=\hat{J}^{2}+\frac{2c_{1}}{1+\cos\theta}+
\frac{2c_{2}}{1-\cos\theta}.
\end{eqnarray*}
Here $\hat{J}^{2}$ is the square of the angular momentum~\cite{Z}
\begin{eqnarray*}
\hat{\bf J}=\frac{1}{\hbar}\left[{\bf r}\times \left({\hat{\bf
p}}+\frac{e}{c}{\bf A}\right)\right] - s\frac{{\bf r}}{r},
\end{eqnarray*}
$\hat{J_{z}}=-\left(s+i\partial/\partial\varphi\right)$ its
$z$-component and $\hat{J_{z}}\psi^{(s)}=m\psi^{(s)}$.

In the parabolic coordinates
\begin{eqnarray} x = \sqrt{\xi \eta}\,\cos\varphi, \qquad y = \sqrt{\xi
\eta}\,\sin\varphi,\qquad z = \frac{1}{2}(\xi - \eta), \qquad
\xi,\eta \in [0, \infty), \qquad \varphi \in [0, 2\pi)
\label{parabolic} \end{eqnarray}
the solution of the equation (\ref{schr}) has the following
form~\cite{mard1}
\begin{eqnarray}
\psi_{n_1n_2m}^{(s)}(\xi,\eta,\varphi;\delta_1,\delta_2) =
\sqrt{2}\varepsilon^{2}\Phi_{n_1m_1}(\xi)
\Phi_{n_2m_2}(\eta)\,\frac{e^{i(m+s)\varphi}}{\sqrt{2\pi}},
\label{parwave1}
\end{eqnarray}
where
\begin{eqnarray*}
\Phi_{n_im_i}(x) = \frac{1}{\Gamma(m_i+1)}
\sqrt{\frac{\Gamma(n_i+m_i+1)}{(n_i)!}}\,\,e^{-\frac{\varepsilon
x}{2}}\,\,(\varepsilon x)^{\frac{m_i}{2}}\,\,F(-n_i; m_i+1;
\varepsilon x).
\end{eqnarray*}
The parabolic quantum numbers $n_1$ and $n_2$ are connected with
the principal quantum number $n$ as follows:
\begin{eqnarray*}
n= n_1 + n_2 + \frac{|m-s|+|m+s|}{2}+1. \label{parqnumb1}
\end{eqnarray*}
It is mentioned~\cite{mard1} that the parabolic basis
(\ref{parwave1}) of the generalized MIC-Kepler system is the
eigenfunction of commuting operators ${\hat J}_z$ and
\begin{eqnarray*}
\hat{X} =\hat{I_z} + \frac{\mu}{\hbar^2}
\left[\lambda_1\frac{r-z}{r(r+z)} -
\lambda_2\frac{r+z}{r(r-z)}\right] \label{intc1}
\end{eqnarray*}
where $\hat{I_z}$ is the $z$ component of the analog of the
Runge-Lenz vector
\begin{eqnarray*}
\hat{\bf I}=\frac{1}{2\sqrt{\mu}}\left[{\hat{\bf J}}\times
\left({\hat{\bf p}}+\frac{e}{c}{\bf A}\right)+ \left({\hat{\bf
p}}+\frac{e}{c}{\bf A}\right)\times{\hat{\bf J}}\right]
+\frac{e^2}{\hbar\sqrt{\mu}}\frac{{\bf r}}{r} \label{runge}
\end{eqnarray*}
and
\begin{eqnarray*}
{\hat X}\psi_{n_1n_2m}^{(s)}(\xi,\eta,\varphi;\delta_1,\delta_2)=
\frac{\hbar \varepsilon}{\sqrt{\mu}}\left(n_1-n_2+m_- +
\frac{\delta_1-\delta_2}{2}\right)
\psi_{n_1n_2m}^{(s)}(\xi,\eta,\varphi;\delta_1,\delta_2).
\label{spectr2}
\end{eqnarray*}

Finally, it is mentioned also that the interbasis expansion of the
parabolic basis over the spherical one has the form~\cite{mard1}
\bea \psi_{n_1n_2m}^{(s)}(\xi,\eta,\varphi;\delta_1,\delta_2) =
\sum_{j=m_+}^{n-1}\,W^{j}_{n_{1}n_{2}ms}\left(\delta_{1},
\delta_{2}\right)\,
\psi_{njm}^{(s)}\left(r,\theta,\varphi;\delta_{1},\delta_{2}\right),
\label{inter}\eea
where
\bea W^{j}_{n_{1}n_{2}ms}\left(\delta_{1}, \delta_{1}\right) =
(-1)^{n_1}\,C^{j+\frac{\delta_{1}+
\delta_{2}}{2},\,\frac{m_1+m_2}{2}}_{\frac{n+m_-
+\delta_2-1}{2},\,\frac{m_2+n_2-n_1}{2};\,\frac{n-m_-
+\delta_1-1}{2},\,\frac{m_1+n_1-n_2}{2}}. \label{W}\eea
Equation (\ref{W}) proves that the coefficients for the expansion
of the parabolic basis in terms of the spherical basis are nothing
but the analytical continuation, for real values of their
arguments, of the $SU(2)$ Clebsch-Gordan coefficients.

Let us demonstrate that if in equation (\ref{schr}) we make the
changes
\begin{eqnarray}
\psi^{(s)}({\bf r}) \to \psi({\bf r},\gamma) = \psi^{(s)}({\bf r})
\frac{e^{is(\gamma-\varphi)}}{\sqrt{4\pi}}, \qquad s \to
-i\frac{\partial}{\partial \gamma},\qquad where \quad \gamma \in
[0, 4\pi), \label{transform}
\end{eqnarray}
it will transform into the Schr\"odinger equation for a
four-dimensional double singular oscillator.

Equation (\ref{schr}) in the spherical coordinates is of the form
\begin{eqnarray}
\frac{1}{r^2}\frac{\partial}{\partial r}\left( r^2\frac{\partial
\psi^{(s)}}{\partial r}\right) + \frac{1}{r^2}\left[\frac{1}{\sin
\theta}\frac{\partial}{\partial \theta}\left( \sin \theta
\frac{\partial \psi^{(s)}}{\partial \theta}\right) +
\frac{1}{\sin^2\theta}\frac{\partial^2\psi^{(s)}} {\partial
\varphi^2}\right] - \frac{2is}{r^2(1-\cos\theta)}\frac{\partial
\psi^{(s)}}{\partial \varphi} - \nonumber \\ \label{3.7.1} \\
- \frac{2s^2}{r^2(1-\cos\theta)}\psi + \frac{2\mu}{\hbar^2}\left[E
+ \frac{e^2}{r}- \frac{\lambda_1}{r^2(1+\cos\theta)}
-\frac{\lambda_2}{r^2(1-\cos\theta)}\right]\psi^{(s)} = 0.
\nonumber
\end{eqnarray}
From (\ref{transform}) and (\ref{3.7.1}) we have
\begin{eqnarray}
\left[\frac{1}{r^2}\frac{\partial}{\partial r}\left(r^2
\frac{\partial}{\partial r}\right) - \frac{{\hat {\bf L}
}^2}{r^2}\right] \psi + \frac{2\mu}{\hbar^2}\left[E +
\frac{e^2}{r}- \frac{\lambda_1}{r^2(1+\cos\theta)}
-\frac{\lambda_2}{r^2(1-\cos\theta)}\right] \psi = 0,
\label{3.7.2}
\end{eqnarray}
where
\begin{eqnarray*}
{\hat {\bf L}}^2 = -\left[\frac{1}{\sin
\beta}\frac{\partial}{\partial \beta} \left(\sin
\beta\frac{\partial}{\partial \beta}\right) +
\frac{1}{\sin^2\beta}\left(\frac{\partial^2}{\partial \alpha^2} -
2\cos\beta\frac{\partial^2}{\partial \alpha \partial \gamma}+
\frac{\partial^2}{\partial \gamma^2}\right)\right].
\end{eqnarray*}
Here we change the notation: $\beta =\theta$ and $\alpha=\varphi$.
If we now pass from the coordinates $r, \alpha, \beta, \gamma$ to
the coordinates
\begin{eqnarray}
u_0 + iu_1 = u\cos{\frac{\beta}{2}} e^{i\frac{\alpha +
\gamma}{2}}, \qquad u_2 + iu_3 = u\sin{\frac{\beta}{2}}
e^{i\frac{\alpha - \gamma}{2}} \label{3.7.3}
\end{eqnarray}
with $u^2 = r$, and take into account that
\begin{eqnarray*}
\frac{{\partial}^2}{\partial u_{\mu}^2} =
\frac{1}{u^3}\frac{\partial}{\partial u}
\left(u^3\frac{\partial}{\partial u}\right) - \frac{4}{u^2}{\hat
{\bf L}}^2, \qquad  \mu =0,1,2,3 \label{3.7.4}
\end{eqnarray*}
and introduce the notations
\begin{eqnarray*}
\epsilon = 4e^2, \qquad E = -\frac{\mu_0 \omega^2}{8}, \qquad
c_i=2\lambda_i \qquad (i=1,2) \label{3.7.5}
\end{eqnarray*}
equation (\ref{3.7.2}) will turn into the Schr\"odinger equation
for a four-dimensional double singular oscillator
\begin{eqnarray}
\left[\frac{{\partial}^2}{\partial u_{\mu}^2} +
\frac{2\mu}{\hbar^2}\left(\epsilon - \frac{\mu\omega^2
u^2}{2}-\frac{c_1}{u_0^2+u_1^2}-
\frac{c_2}{u_2^2+u_3^2}\right)\right] \psi({\bf u}) = 0,
\label{5.5.9}
\end{eqnarray}
whose energy spectrum is given by the formula
\begin{eqnarray*}
\epsilon = \hbar \omega\left(N+\delta_1+\delta_2+2\right).
\end{eqnarray*}

Using formulae (\ref{spherical}) and (\ref{3.7.3}) and considering
that $r=u^2,\,\theta = \beta,\,\varphi = \alpha$, one can easily
show that
\begin{eqnarray*}
x &=& 2(u_0u_2 - u_1u_3)\,, \\ [2mm] y &=& 2(u_0u_3 + u_1u_2)\,,
\\ [2mm] z &=& u_0^2 + u_1^2 - u_2^2 - u_3^2\,
\\ [2mm] \gamma &=& \frac{i}{2}\ln{\frac{(u_0-iu_1)(u_2+iu_3)}
{(u_0+iu_1)(u_2-iu_3)}}\,.
\end{eqnarray*}
The first three lines are the transformation ${\rm I \!R}^4 \to
{\rm I \!R}^3$ suggested by Kustaanheimo and Stiefel for the
regularization of the equations of celestial mechanics~\cite{KS}.
Later, this transformation found other applications, as
well~\cite{Barut,KN}. This transformation supplemented with the
coordinate $\gamma$ (generalized Kustaanheimo-Stiefel
transformation) was used for the "synthesis" of the charge-dyon
system from the four-dimensional isotropic oscillator~\cite{NTA}.

Introducing the double polar coordinates
\begin{equation}
u_0+i u_1= \rho_1\,e^{i\varphi_1}, \quad u_2+i u_3= \rho_2\,
e^{i\varphi_2}, \qquad {\rm where} \quad \rho_1,\rho_2 \in
[0,\infty), \quad \varphi_1,\varphi_2 \in [0,2\pi). \label{3.8.3}
\end{equation}
From the formulae (\ref{spherical}), (\ref{parabolic}),
(\ref{3.7.3}) and (\ref{3.8.3}) we get the relations
\begin{eqnarray*}
\xi= 2\rho_1^2, \qquad \eta = \rho_2^2, \qquad \varphi =\varphi_1
+ \varphi_2, \qquad \gamma=\varphi_1 - \varphi_2
\end{eqnarray*}
which lead to the formulae
\begin{eqnarray*}
\psi_{NLMM'}\left(u,\alpha,\beta,\gamma\right)=
4\left(n+\frac{\delta_1 +\delta_2}{2}\right)\sqrt{a}
\delta_{n,\frac{N}{2}+1}\delta_{jL}\delta_{m M}\delta_{sM'}
\psi_{njms}\left(r,\theta,\varphi,\gamma\right),
\end{eqnarray*}
\begin{eqnarray*}
\psi_{N_1N_2M_1M_2}\left(\rho_1,\rho_2,\varphi_1,\varphi_2\right)=
4\left(n+\frac{\delta_1 +\delta_2}{2}\right)\sqrt{a} \times\\
[2mm] \times
\delta_{n_1,N_1}\delta_{n_2,N_2}\delta_{m,\frac{M_1+M_2}{2}}\delta_{s,
\frac{M_1-M_2}{2}}
\psi_{njms}\left(\rho_1,\rho_2,\varphi_1,\varphi_2\right)
\end{eqnarray*}
generalizing the earlier results~\cite{M1}.

Now we are able to write the expansion (\ref{inter})
\bea
\psi_{N_1N_2M_1M_2}\left(\rho_1,\rho_2,\varphi_1,\varphi_2\right)
= \sum_{L=L_{min}}^{N/2}\,W^{NLMM'}_{N_{1}N_{2}M_1M_2} \,
\psi_{NLMM'}\left(u,\alpha,\beta,\gamma\right)_,
\label{inter1}\eea
\begin{eqnarray*} W^{NLMM'}_{N_{1}N_{2}M_1M_2} = e^{i\pi\Phi}C_{a_0, \alpha_0;
b_0, \beta_0}^{c_0, \gamma_0},
\end{eqnarray*}
where
\begin{eqnarray*}
a_0=\frac{N_1+N_2+|M-M'|+\delta_2}{2}, \quad
\alpha_0=\frac{N_2-N_1+|M-M'|+\delta_2}{2}, \\ [2mm]b_0
=\frac{N_1+N_2+|M+M'|+\delta_1}{2}, \quad
\beta_0=\frac{N_1-N_2+|M+M'|+\delta_1}{2}, \\ [2mm]c_0 =
L+\frac{\delta_1+\delta_2}{2}, \quad \gamma_0 =
\frac{|M+M'|+|M-M'|+\delta_1+\delta_2}{2}.
\end{eqnarray*}
The lower limit of summation in (\ref{inter1}) and quantity $\Phi$
are given by the expressions
\begin{eqnarray*}
L_{min}=\frac{1}{2}\left(|M+M'|-|M-M'|\right), \quad
\Phi=N_1+\frac{1}{2}\left(M-M'+|M-M'|\right).
\end{eqnarray*}

\vspace{5mm}

{\large Acknowledgements.} I would like to thank A. Nersessian and
G.Pogosyan for useful discussions. The work is carried out with
the support of NFSAT-CRDF grant ARPI-3228-YE-04.

\end{document}